\documentclass[aps,prb,reprint, superscriptaddress]{revtex4-1} 

\usepackage{hyperref, natbib}
\usepackage{graphicx} 
\DeclareGraphicsExtensions{.pdf,.png,.jpg}

\usepackage{amsmath}
\setcitestyle{super}

\usepackage[normalem]{ulem} 
\usepackage{color}

\usepackage{gensymb}

\begin{document}



\title{Magnetic order and energy-scale hierarchy in artificial spin ice}


\author{Henry Stopfel}%
\affiliation{%
 Department of Physics and Astronomy, Uppsala University, Box 516, SE-75120, Uppsala, Sweden
}%
\author{Erik \"{O}stman}
\affiliation{%
 Department of Physics and Astronomy, Uppsala University, Box 516, SE-75120, Uppsala, Sweden
}%
\author{Ioan-Augustin Chioar}%
\affiliation{%
 Department of Physics and Astronomy, Uppsala University, Box 516, SE-75120, Uppsala, Sweden
}%
\author{David Greving}
\affiliation{
 Department of Physics, University of Warwick, Coventry, United Kingdom
}%
\author{Unnar~B.~Arnalds}%
\affiliation{%
 Science Institute, University of Iceland, Reykjavik, Iceland
}%
\author{Thomas P. A. Hase}
\affiliation{
 Department of Physics, University of Warwick, Coventry, United Kingdom
}%
\author{Aaron Stein}%
\affiliation{%
 Center for Functional Nanomaterials, Brookhaven National Laboratory, Upton, New York 11973, USA
}%
\author{Bj\"{o}rgvin Hj\"{o}rvarsson}%
\affiliation{%
 Department of Physics and Astronomy, Uppsala University, Box 516, SE-75120, Uppsala, Sweden
}%
\author{Vassilios Kapaklis}%
\affiliation{%
 Department of Physics and Astronomy, Uppsala University, Box 516, SE-75120, Uppsala, Sweden
}%

\date{\today}
\maketitle

{\bf In order to explain and predict the properties of many physical systems, it is essential to understand the interplay of different energy-scales. Here we present investigations of the magnetic order in thermalised artificial spin ice structures, with different activation energies of the interacting Ising-like elements. We image the thermally equilibrated magnetic states of the nano-structures using synchrotron-based magnetic microscopy. By comparing results obtained from structures with one or two different activation energies, we demonstrate a clear impact on the resulting magnetic order.  The differences are obtained by the analysis of the magnetic spin structure factors, in which the role of the activation energies is manifested by distinct short-range order. This demonstrates that artificial spin systems can serve as model systems, allowing the definition of energy-scales by geometrical design and providing the backdrop for understanding their interplay.
}

\section{Introduction}

Initially introduced as a playground for the experimental investigation of magnetic frustration effects\cite{Wang2006Nat}, the area of artificial spin ices  has evolved into a vibrant field of research. Using designed structures, collective order and dynamics can be controlled and directly visualized by using, for example, nano-characterization techniques\cite{Nisoli2013RevModPhys}. This approach has motivated experimental investigations of celebrated classical spin models\cite{Syozi_1951,Lieb1967PRL,Lieb1967PhysRev}, some of which display non-conventional and exotic magnetic phases\cite{Zhang2013Nat,Canals_NatComm_2016,Perrin_Nature_2016, Nisoli_NatPhys_2017,Erik_ArXiV_2017}. Furthermore, given the large freedom in design, new topologies that promote non-conventional emergent order can be manufactured and investigated in thermally-active artificial spin ice structures, allowing the study of their complex energetic manifolds in a superparamagnetic regime \cite{Morgan2010NatPhys,Arnalds2012APL, Farhan2013NatPhys, Farhan2013PRL,Kapaklis2012NJP, Kapaklis2014NatNano, mSH_NatPhys_2018}. A variety of new artificial spin ice systems have been proposed by Morrison \textit{et al.}\cite{Morrison2013NJP}, with particular attention given to the so-called Shakti lattice, a variation of the square ice geometry.  In this lattice, 25\% of the square ice lattice elements have been removed and certain pairs of islands merged as depicted in Fig.~\ref{fig1}. 
Theoretical and numerical studies have predicted an emergent magnetic order in the Shakti geometry\cite{Chern2013PRL} which retrieves the features of the six-vertex model\cite{Lieb1967PRL} on a larger length-scale. This behavior has been demonstrated experimentally by Gilbert \textit{et al.}\cite{Gilbert2014NatPhys} for the Shakti lattice (SH), as well as for a modified Shakti lattice (mSH). 

Given the mesoscopic nature of the structures, different activation energies are expected for the long (T$_{\rm{B1}}$) and short islands (T$_{\rm{B2}}$), resulting in two blocking temperatures of the elements (T$_{\rm{B2}} < T_{\rm{B1}}$). The presence of two activation energies is therefore expected to influence the eventual magnetic correlations and order in the lattice, since the larger elements could act as a source of quenched disorder. Such disorder would hinder the development of any medium- or long-range magnetic order of the smaller islands, as these become arrested at lower temperatures (T$_{\rm{B2}}$). Whilst the vertex statistics for both the Shakti lattices have been experimentally investigated for a range of inter-island coupling strengths\cite{Gilbert2014NatPhys}, the thermally driven kinetics in the modified Shakti lattice were only recently reported\cite{mSH_NatPhys_2018}. Consequently, the impact of the hierarchy of the activation energies on the emergent order has not been addressed to date. In this work, we undertake this task for the particular case of the Shakti and the modified Shakti geometries. Using thermally-active elements, we characterize and compare their final magnetic configurations using identical experimental protocols, highlighting the impact of different activation energies on the obtained magnetic correlations and order.

\section{Experimental details}

\subsection{Sample preparation}

\begin{figure*}[t!]
	\includegraphics[width=\textwidth]{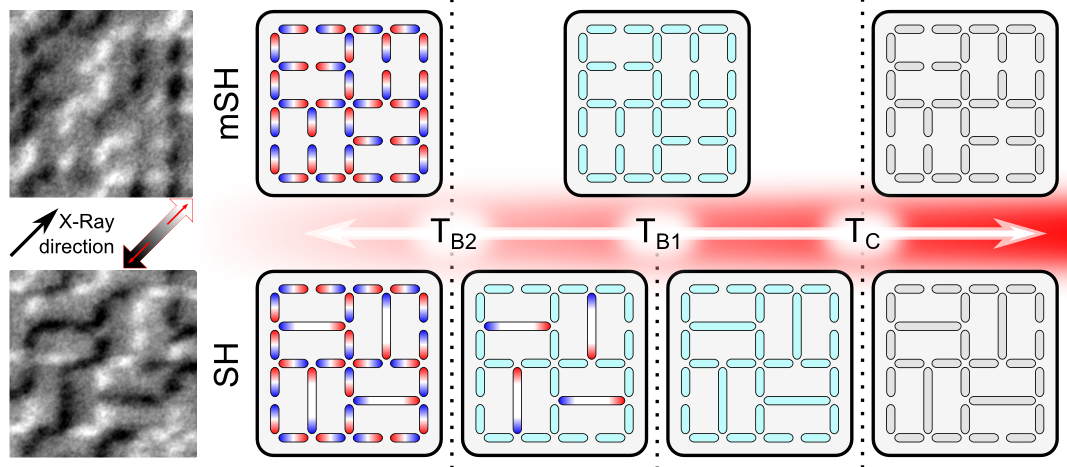}
	\caption{A schematic representation of the Shakti (SH) and modified-Shakti (mSH) lattices along with magnetic states representative of each temperature regime, generally defined by the Curie (T$_{\rm{c}}$) and the blocking temperatures (T$_{\rm{B1/B2}}$). Elements in their paramagnetic state are gray, superparamagnetic elements are light-blue and the frozen elements are represented with a magnetic north (blue) and south (red) pole, reflecting their final magnetic orientation. The gray-scale images to the left are representative PEEM-XMCD snapshots obtained at 65~K, the lowest temperature reached during the cooling.}
	\label{fig1}
\end{figure*}

The samples were produced by a post-growth patterning process, applied to a thin film of $\delta$-doped Palladium (Iron) as described in P\"arnaste \textit{et al.}\cite{Parnaste2007}.  A thick bottom layer of Pd (40~nm) is followed by 2.0 mono layers of Fe, defining the nominal Curie-temperature, T$_{\rm{C}}=400$~K\cite{Parnaste2007}, as well as the thermally active temperature range for the $\delta$-doped Pd(Fe). The post-patterning was carried out at the Center for Functional Nanomaterials (CFN), Brookhaven National Laboratory in Upton, New York. A positive high resolution e-Beam resist (ZEP520A) was employed to create a Chromium mask of the nano-structures using e-Beam lithography. The magnetic structures were formed by Argon milling. All investigated structures were produced on the same substrate from the same layer, ensuring identical intrinsic material properties as well as the same thermal history for all the investigated structures. Each array had a spatial dimension of 200 $\times$ 200~$\mu$m$^2$. The lengths of the large and small islands, were chosen to be 1050~nm and 450~nm, respectively, while the width was kept at 150~nm. The lattice spacing between two parallel neighboring short islands was set to 600~nm. 

\subsection{Thermal protocol}

Similar to previous studies\cite{Kapaklis2012NJP,Kapaklis2014NatNano}, the thermal protocol involves a gradual cooling from a superparamagnetic state of the elements, towards a completely arrested state at the lowest temperatures. This process is schematically represented in Fig.~\ref{fig1} for both the Shakti and the modified Shakti lattices. Notice the two distinctive stages of the superparamagnetic regime for the Shakti lattice because to the presence of elements with different intrinsic activation energies.

When the temperature is lowered below the Curie temperature (T$_{\rm{c}}$),  the elements become magnetic and can be considered as mesospins. The stray field of these islands is resulting in a temperature-dependent magnetostatic coupling. The coupling between the elements has a direct impact on the involved activation energies, biasing the otherwise equally-favoured two magnetic states of the islands. On the lattice scale, this effectively results in a dynamic distribution of blocking temperatures around the intrinsic energy barriers, a distribution whose width depends on the strength of the inter-island interactions. 

\subsection{Interaction strength and their impact}

There are three special cases to be considered: the strong, intermediate and weak coupling limits. In the strong coupling limit, the intrinsic activation energy of the mesospins is much smaller than the inter-island interactions. Consequently, the emergent order will not depend on the differences in the activation energies, but significant effects on the dynamic response cannot be ruled out. In the intermediate coupling limit, the strength of the inter-island couplings is comparable to the intrinsic activation energy of the mesospins. This will give rise to a complex interplay, in which the relation between the activation and interaction energies will dictate both the order and dynamics of the magnetic phases. The individual activation energies depend on the dynamically developing magnetic configurations, for both long- and short-mesospins. In other words, these couplings can yield a strong overlap in the two blocking temperature distributions of the short and long islands. Therefore, the configurations of the long- and the short-mesospins will exhibit different degrees of order, all depending on the relation of the activation and interaction energies. 
Conversely in the weak coupling limit, the intrinsic activation energies of the long-mesospins are much larger than the inter-island interactions. Around the intrinsic blocking temperature of the long islands (T$_{\rm{B1}}$), the available thermal energy is higher than the energy difference given by any magnetization reversal of the short islands. 
Therefore the lattice as a whole will be in a disordered state when the long-mesospins freeze (T$_{\rm{B1}}$)  and consequently the magnetic orientation of the long elements will be randomly distributed, with no short- or long-range correlations. However, at the blocking temperature for the short-mesospins (T$_{\rm{B2}}$) the inter-island coupling strength can be comparable to their intrinsic activation energies, potentially resulting in a short-range order below T$_{\rm{B2}}$. Thus, having two distinct activation energies is expected to have a significant impact on the emergent order, for both the intermediate and the weak coupling limits. By the same token, the presence of two or more activation energies is expected to have a marginal impact in the strong coupling limit. Here, we focus on the weak coupling limit and, as previously stated, we compare the obtained magnetic order in SH  and mSH lattices, ascertaining the impact of the hierarchical relation of inter-element interactions and the activation energies involved. 


\subsection{Determination of the magnetization direction}

The magnetic states were determined as a function of temperature using Photo Emission Electron Microscopy (PEEM) capturing the X-ray Magnetic Circular Dichroism (XMCD) contrast. The experimental studies were performed at the 11.0.1 PEEM3 beamline at the Advanced Light Source, in Berkeley, California, USA. For the SH and mSH lattices the islands were oriented 45$\degree$ with respect to the incoming X-ray beam, to enable identification of the magnetization direction of all elements in the lattices. For these lattices multiple, partially  overlapping, PEEM-XMCD images were acquired and stitched together to form one large image containing over 4000 mesospins for each of the investigated lattices, see Supplementary Materials Fig.~2. The images used for determining the magnetic states were taken at 65~K, far below the Curie-temperature and the blocking temperatures of the short- and long-mesospins. The sampling time ($t_{s} = 300~s$) required for obtaining a PEEM-XMCD image defines a time window linked to the thermal stability of the mesospins: At the lowest temperatures, all magnetic states are stable (frozen) and the lattice can be imaged for practically arbitrary long times, yielding the same result. However, if the reversal times are much smaller than $t_{s}$, no magnetic contrast is obtained. This allowed us to ask at which temperature the spins have reversed their magnetization direction within a given time window, thereby providing an estimation on the involved energy scales. 

\section{Results}

\subsection{Activation energies of the mesospins}

Magnetic interactions can affect the inferred activation energies in a profound way.  
We therefore fabricated two additional lattices consisting of long-  and short-mesospins, 1110~nm and 450~nm respectively, in which the distance was large enough to allow us to ignore any effects arising from the interactions between the elements (see Supplementary Materials). The islands were oriented parallel with respect to the incoming X-ray beam, to determine the magnetization direction of the mesospins.

Following the line of argumentation above, we determined that 50\% of the short islands reverse their magnetization direction within 300~s  at  112~K,  while a temperature of 149~K is required for obtaining the same condition for the long islands. 
The determination of the activation energies is model dependent. 
Using an expression for the probability of a magnetic element not having altered its magnetization direction after a time period $t_s$\cite{Wernsdorfer1997PRL, Kapaklis2014NatNano}, as described in the Supplementary Materials, yields a ratio for the activation energies of 0.49(4). This approach provides support for well separated energy-scales, which will be assumed to be the case in the remainder of this communication.


\begin{figure}[b!]
	\includegraphics[width=\columnwidth]{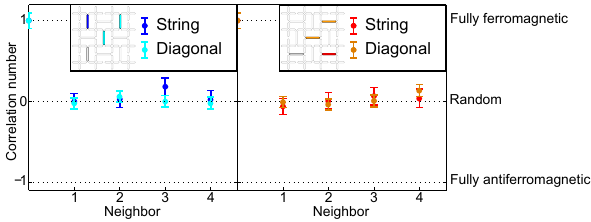}
	\caption{Long-mesospin correlations in the Shakti lattice. The network-averaged magnetic correlations for long-mesospins spatially oriented vertically/horizontally are shown in the left/right graph. In blue/red the correlations of long-mesospins which are oriented in a string and in light-blue/orange the correlations of the long-mesospins diagonally oriented are represented, as indicated in the inset. All correlations show that already the first neighbor is randomly oriented and therefore the magnetic orientations of the long-mesospins are not correlated. Error bars represent one standard deviation.}
	\label{fig2}
\end{figure}

\subsection{Correlations of the long-mesospins in the Shakti lattice} 

The correlations between the long-mesospins were determined from the PEEM-XMCD images recorded at 65~K, after cooling the samples from room temperature with a  rate of 1~K/minute. The results from analysis along different directions in the SH lattice are illustrated in Fig.~\ref{fig2}. As seen in the figure, no correlations between long-mesospins are observed. 
The cooling of the sample below the blocking temperature of the short-mesospins does not result in any observed order for the long-mesospins. Because of this, the effective interaction mediated by the short-mesospins must be much smaller than the activation energy of the long-mesospins.

\begin{figure}[t!]
\includegraphics[width=\columnwidth]{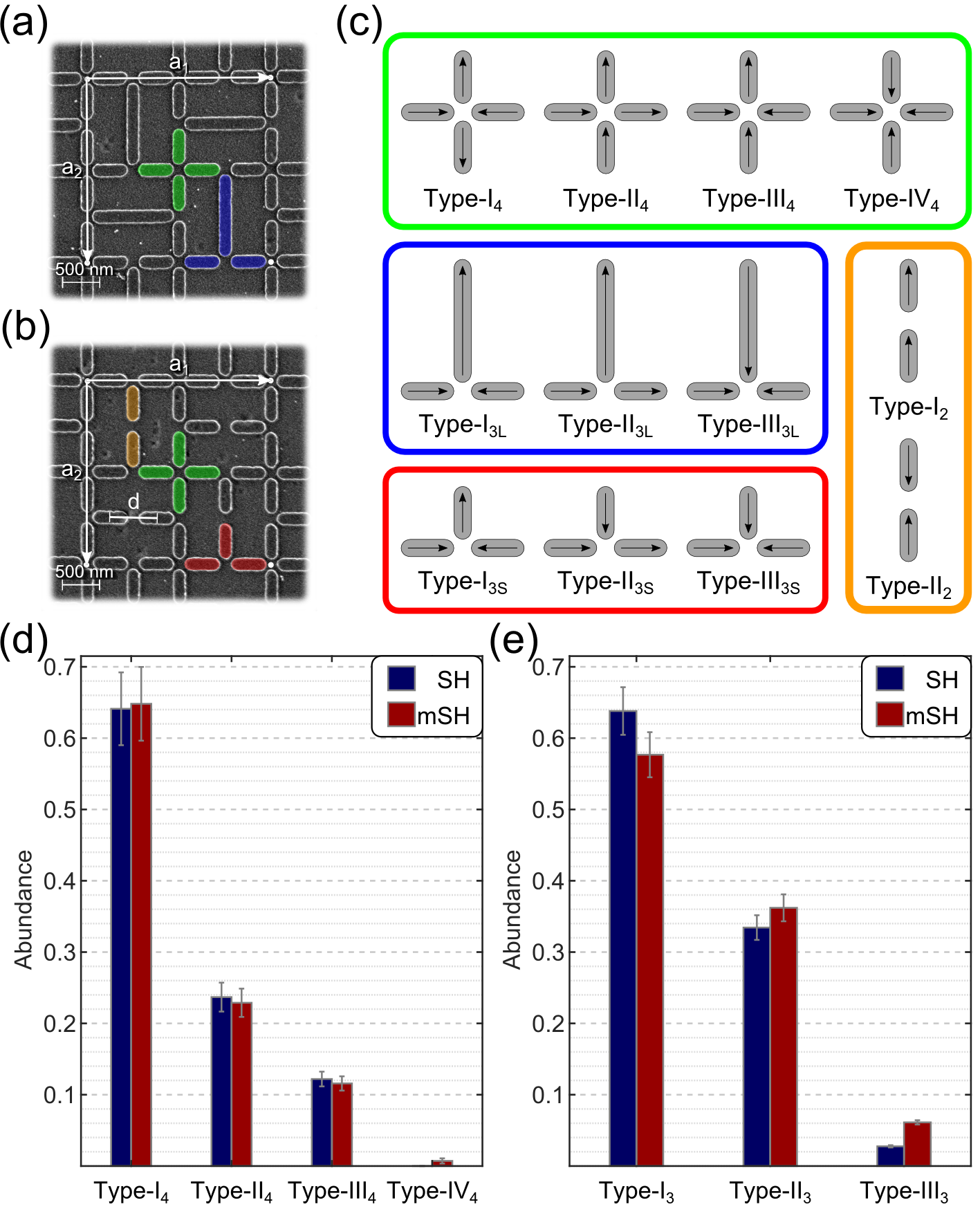}
\caption{Vertex type definition and statistics for SH and mSH lattices. SEM image of the SH ({\bf a}) and mSH ({\bf b}) lattice. The different vertex types are indicated by their respective colors. The real space base vectors for the Shakti lattice are defined along the x- and y-direction and each have a length of four times the lattice spacing ($d = 600~\rm{nm}$). ({\bf c}),~Definition of the vertex states for all different vertex types. While both lattices share the vertex type with a coordination number of four (green), the mSH lattice contains vertices with coordination number three (red) and two (orange). In the SH lattice the three-fold coordinated vertices (blue) include a long-mesospin. 
The four- ({\bf d}) and three-fold ({\bf e}) coordinated vertex statistics for SH (dark-blue) and mSH (dark-red) lattices show no significant differences. Error bars represent one standard deviation.}
\label{fig3}
\end{figure}

\subsection{Magnetic order on the vertex-level}

Having established the randomness of the magnetic states of the long-mesospins in the Shakti lattice, we now turn our attention to the overall magnetic order established within the SH and mSH lattices. For square ice related geometries, a commonly-employed tool for characterizing the obtained magnetic state is the determination of populations of the various types of vertex configurations. The different vertex types for both the SH and the mSH lattices are illustrated in Fig.~\ref{fig3}({\bf a})-({\bf c}).
In the nomenclature used here, a roman numeral denotes the vertex type along with a subscript representing the number of mesospins forming the vertex. As an example, I$_{4}$ represents four elements forming a Type-I vertex. The mSH lattice has an additional degree of freedom as compared to the SH lattice, arising from the two-fold coordinated vertex having two possible states: Type-I$_{2}$ and Type-II$_{2}$.

The vertex statistics obtained for the three- and four-fold coordinated vertices are presented in Fig.~\ref{fig3}({\bf d}) and \ref{fig3}({\bf e}), respectively. Comparing the abundance of the different states with published data from Gilbert {\it et al.}\cite{Gilbert2014NatPhys}, we can again confirm that our results are representative for the weak coupling regime, as $\it{e.g.}$ the previous works on ASI structures\cite{Arnalds2012APL, Kapaklis2014NatNano}. Notice that these vertex statistics do not display a significant difference between the common vertex states of the SH and mSH lattices, similar to previous reports\cite{Gilbert2014NatPhys}. This implies that the final vertex populations are independent of the presence of the long islands in the Shakti geometry and the related energy hierarchy. However, the vertex statistics yield only a partial description of the final magnetic state and are insensitive to magnetic order established over spatial extensions larger than the vertex size.


\subsection{Beyond the vertex-level}
We use the calculated spin structure factor (SSF) to determine correlations beyond nearest neighbours\cite{Canals_NatComm_2016, Perrin_Nature_2016, Erik_ArXiV_2017, Farhan2017NatCom}. This invokes the use of a Fourier transformation of pairwise correlations of the magnetic states in the following form:

\begin{equation}
	I(\mathbf{q})=\frac{1}{N^2}\sum_{(i,j=1)}^{N}\mathbf{S}^{\perp}_i\cdot \mathbf{S}^{\perp}_j \cdot e^{i\mathbf{q}\cdot(\mathbf{r}_i-\mathbf{r}_j)}
	\label{eq:SSF}
\end{equation}

\begin{figure}[b!]
	\includegraphics[width=\columnwidth]{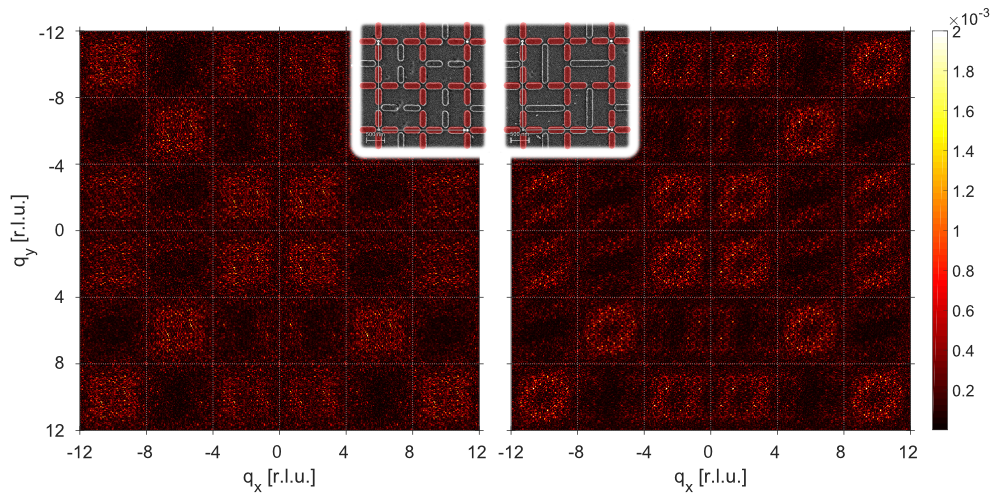}
	\caption{Spin structure factor maps computed for the four-fold coordinated vertex sub-lattices of the mSH (left) and the SH (right) lattices. The real space lattice vectors as well as the sub-set of four-fold coordinated vertices (red islands) are indicated in the insets.}
	\label{SSF}
\end{figure}

\begin{figure*}[t!]
\includegraphics[width=\textwidth]{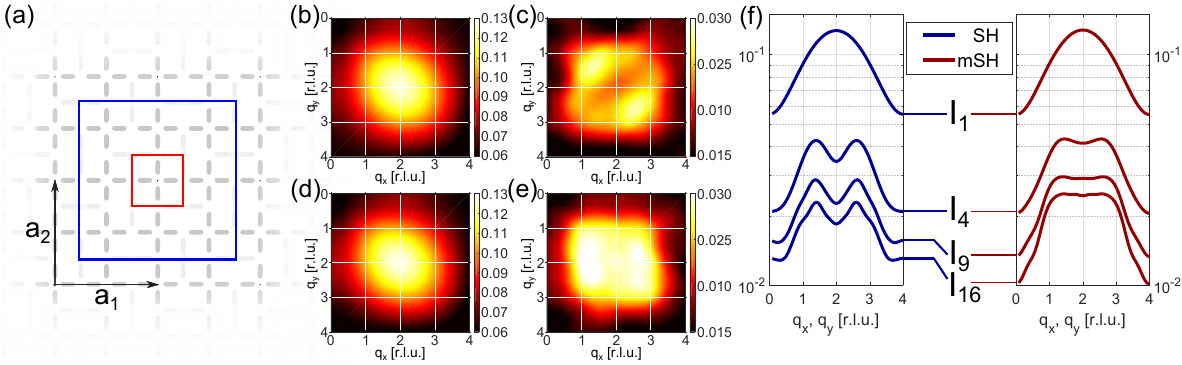}
\caption{Spatially-limited spin structure factor as indicator for short-range order. ({\bf a}), Illustration of the mesospin sub-set used in the SSF calculations. The real space lattice vectors ($\mathbf{a}_{1},\mathbf{a}_{2}$) which are used for the reciprocal lattice units (r.l.u.) are also shown. The real space input window size for the averaged SSF $I_1$ (red -- SSF ({\bf b/d}) ) and $I_9$ (blue -- SSF ({\bf c/e}) ) are highlighted. {\bf b-e}, The averaged SSF maps for SH ({\bf b/c}) and mSH ({\bf d/e}) lattices, with reduced real space input size according to equation \eqref{SSF_avg}. ({\bf f}), Diagonal line-profiles for the SSF of different input sizes for the SH (dark-blue) and mSH (dark-red) lattices. The modulation of the intensity in the SSF maps at (q$_x$, q$_y$) = (2, 2) [r.l.u.] is clearly distinguishable for the case of the SH lattice and an indication of a different order for these length-scales. }
\label{fig5}
\end{figure*}

\noindent where $I(\mathbf{q})$ is the intensity for a given scattering vector $\mathbf{q}$ in reciprocal space, $N$ denotes the number of spins over which the SSF is calculated, $\mathbf{S}^{\perp}_{i}$ the spin component perpendicular to $\mathbf{q}$ and $\mathbf{r}_{i}$ is the position of the corresponding spin in the real-space lattice. The SH and the mSH lattices have different lattice points, $\mathbf{r}_{i}$, for the long- and two short-mesospins, the differences in the SSF of the SH and mSH data is dominated by this structural difference. To facilitate a proper comparison between the two cases, only the mesospin sub-set found in both geometries, namely the four-fold coordinated vertices, has been taken into account for the SSF computation. Therefore we calculate the SSF according to equation \eqref{eq:SSF}, in the same manner as is explained in more detail by \"Ostman {\it et al.}\cite{Erik_ArXiV_2017}, but only using the mesospins of the vertices with a coordination number of four. Since this sub-lattice of mesospins is the same in both the SH and mSH lattices, differences in the SSF represent different ordering among these sub-sets of spins. Characteristic SSF maps for the SH and mSH networks are presented in Fig.~\ref{SSF}.\\ 

Although neither of the two maps presents specific features, such as pinch points or Bragg peaks, their diffuse signals systematically exhibit clear differences. This is indicative of differences in the global spin order, which have not been captured by the statistics determined on the vertex-level (presented in Fig.~\ref{fig3}({\bf d}) and \ref{fig3}({\bf e}) ). The striking feature is the presence of a hollow region ($\it{e.g.}$ at the ($\pm$2,$\pm$2) (r.l.u.) points in reciprocal space) in the SH lattice SSF map, with the same region having a more uniform intensity distribution in the structure factor map of the mSH. To understand the origin of this effect and identify the length-scales that are at play, we have computed magnetic structure factors for networks of gradually-increasing size and network-averaged these individual signals for each chosen network-size. Since the sample is large enough to assume ergodic conditions, this size reduction of the sampling window and the corresponding network-averaging, provides statistical information about the magnetic state on the length-scale of the sampling window rather than the high spatial frequencies of the speckle pattern (Fig.~\ref{SSF}) obtained from the specific global microstate.

Hence, we define 
	\begin{equation}
		I_L = \frac{1}{\left({4L}\right) ^2 M}\sum_{m=1}^{M}\sum_{(i,j=1)}^{4L} \mathbf{S}^{\perp}_i\cdot \mathbf{S}^{\perp}_j \cdot e^{i\mathbf{q}\cdot(\mathbf{r}_i-\mathbf{r}_j)}
		\label{SSF_avg}
	\end{equation}
as an incoherent\cite{Incoherent} spatial average over sampling windows with $L$ four-fold coordinated vertices. Each sampling window therefore consists of $4L$ spins (taking into account only the short-mesospins). The SSF $I_L$ is the average of all $M$ sampling windows present in the measured global microstate.

Fig.~\ref{fig5}({\bf b}) and \ref{fig5}({\bf d}) show a comparison of $I_1$, i.e.~the averaged SSF for a single four-fold coordinated vertex (four mesospins) for the SH and mSH lattices. Both are virtually indistinguishable, which is not surprising since $I_1$ is merely a different representation of the vertex statistics shown in Fig.~\ref{fig3}({\bf d}). In other words, as the vertex statistics showed no significant differences, the SSF maps in Fig.~\ref{fig5}({\bf b}) and Fig.~\ref{fig5}({\bf d}) should also be similar, traceable also in a comparison of the top-most diagonal line-profiles in Fig.~\ref{fig5}({\bf f}). Expanding the SSF input window to sixteen mesospins ($I_4$, four four-fold coordinated vertices), a clear difference between the SH and mSH appears, with a more pronounced dip in the intensity at the (2,2) reciprocal space position for the SH lattice case (Fig.~\ref{fig5}({\bf f}) ). Expanding the real space input window to thirty-six mesospins ($I_9$, nine four-fold coordinated vertices), the differences in the SSF maps between the SH (Fig.~\ref{fig5}({\bf c}) ) and mSH (Fig.~\ref{fig5}({\bf e}) ) become even more pronounced (see also the line-profiles in Fig.~\ref{fig5}({\bf f}) ). The line-profiles in Fig.~\ref{fig5}({\bf f}) are diagonal cuts through the respective SSF maps $I_1$, $I_4$, $I_9$ and $I_{16}$ from reciprocal space position (q$_x$, q$_y$) = (0, 0) [r.l.u.] to (q$_x$, q$_y$) = (4, 4) [r.l.u.]. The results from these spatially-limited SSF maps in Fig.~\ref{fig5} are qualitatively similar to the features observed for the SSF of the full microstate in Fig.~\ref{SSF} and capture differences in the magnetic short-range order, exceeding the extension of just one vertex. \\

\begin{figure}[t!]
	\includegraphics[width=\columnwidth]{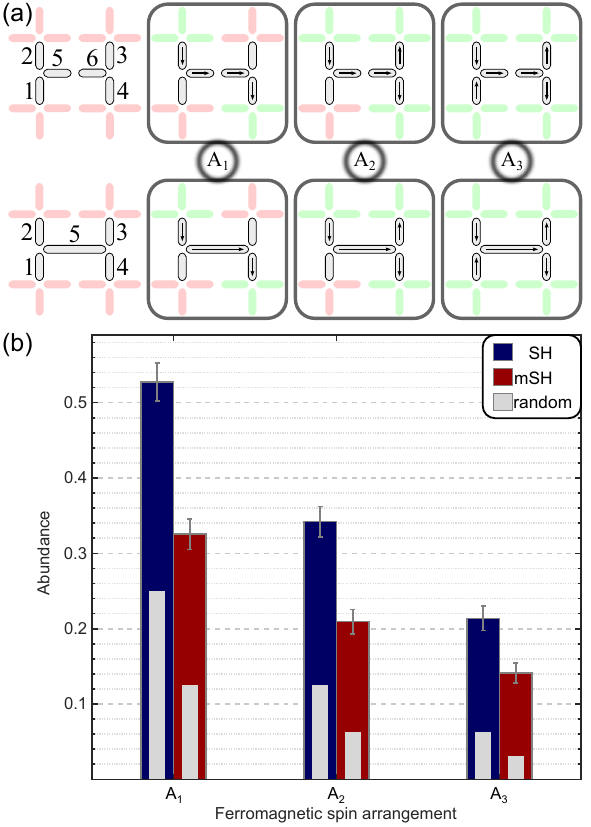}
	\caption{Spin arrangements across the long-mesospins / two-fold coordinated vertices. ({\bf a}), Schematic representation of all mesospins contained within two neighboring three-fold coordinated vertices (gray numbered islands). We classify the arrangements of all these mesospins in three classes, denoted as A$_1$, A$_2$ and A$_3$. The abundance values of A$_1$ and A$_2$ are averages of the four possible spin arrangements in these classes. One representative spin arrangement for each of the three classes is depicted in ({\bf a}). The coupled four-fold coordinated vertices are indicated by the light-green islands. The number of coupled four-fold coordinated vertices across the Type-I$_2$ vertex or a long-mesospin is two for A$_1$, three for A$_2$ and all four for A$_3$. ({\bf b}), The abundance of these three classes is plotted for the SH (dark-blue) and the mSH (dark-red) lattices, the abundance for each spin arrangement in a random mesospin state is represented with a gray inside bar. The abundance revealing the distinct differences in the arrangement of the mesospins around the Type-I$_2$ vertex and is strongly indicating a defined tiling for the four-fold coordinated vertices around the long-mesospins and the build-up of a short-range order. Error bars represent one standard deviation.}
	\label{fig4}
\end{figure}

\subsection{Short-range magnetic order}
We attribute the origin of the features observed in the SSF maps to the long elements in the SH lattice and the difference in activation energies between the short and long islands. To elaborate more on this premise, we have extracted the abundance of various spin arrangements that connect four-fold coordinated vertices across a Type-I$_2$ vertex in the mSH lattice or a long-mesospin in the SH lattice. Here, we focus on three of the different configurations of interest for such spin arrangements, as presented in Fig.~\ref{fig4}({\bf a}). In the three vertex coupling classes, A$_1$, A$_2$ and A$_3$, a total of nine different ferromagnetic spin arrangements are investigated. A detailed analysis of all nine spin arrangements can be found in the Supplementary Materials Fig.~3. In Fig~\ref{fig4}({\bf b}) we report their abundances for both lattices, along with the estimated values given by a random distribution, in the absence of any interaction between the considered mesospins. A$_1$ and A$_2$ are averages of four different spin arrangements (C$_1$-C$_4$ and C$_5$-C$_8$, see Supplementary Materials Fig.~3), which have the same characteristic. In the case of A$_1$, one mesospin (mesospin 1 or 2 in schematic \ref{fig4}({\bf a}) ) is ferromagnetically aligned to the long-mesospin (or the two ferromagnetically aligned mesospins of the two-fold coordinated vertex) and one mesospin on the other side (mesospin 3 or 4 in Fig.~\ref{fig4}({\bf a}) ). For A$_2$ one mesospin on one side is ferromagnetically aligned with two mesospins on the other side of the long-mesospin (or the two ferromagnetically aligned mesospins of the two-fold coordinated vertex). A$_3$ represents the lowest energy configuration for these five (six in the mSH) mesospins and arranges all mesospins (mesospin 1, 2, 3 and 4 in schematic Fig.~\ref{fig4}({\bf a}) ) ferromagnetically with respect to the long-mesospin (or the two ferromagnetically aligned mesospins of the two-fold coordinated vertex).

\section{Discussion}

The spin arrangements in Fig.~\ref{fig4} highlight the development of a stronger correlation between four-fold coordinated vertices across a long-mesospin in the SH lattice than in the mSH, in which case the coupling is mediated via a two-fold coordinated vertex.
In the Shakti ground state manifold as described by Chern {\it et al.}\cite{Chern2013PRL}, the four-fold coordinated vertices are always in their lowest energy state, Type-I$_4$ (see definition Fig.~\ref{fig3}{\bf c}). This state has a two-fold degeneracy, which we denote by Type-I$_4$(A) and Type-I$_4$(B) (see Supplementary Materials Fig. 4({\bf b}) ). A transition from Type-I$_4$(A) to a Type-I$_4$(B) is given by a magnetization reversal of all four mesospin. Two Type-I$_4$ vertices which are ferromagnetically coupled via a Type-I$_2$ or a long-mesospin will always be of opposite states, Type-I$_4$(A) and Type-I$_4$(B). 
Returning to the SSF maps in Fig.~\ref{SSF} and Fig.~\ref{fig5}, the distinct shape in the SSF can be explained by the enhanced effective coupling of Type-I$_4$(A) to Type-I$_4$(B) vertices via the long-mesospins in the SH lattice. 
In other words, although the long-mesospins freeze in a random configuration, they are more effective in transmitting magnetic correlations from one end of a Shakti plaquette to the other, than the two-fold vertex of the mSH lattice. While this is a rather straightforward result for the case of one plaquette, it is \emph{a priori} expected that, on the lattice scale, the prior establishment of a random distribution of long-mesospin states would yield an even more randomized configuration of short islands, a fact that the mSH lattice might circumvent given the absence of these constraints. 

To further discuss our results we would like to compare them with the ones previously reported by Gilbert {\it et al.}\cite{Gilbert2014NatPhys}. For the strong coupling regime, they observed an almost perfect ground-state ordering of the four-fold coordinated vertices for the mSH lattice, while the SH lattice has only about 80\% of these vertices in the lowest energy state, Type-I$_4$. These differences, in the light of our results, can be attributed to the distinct energy-scales. As the long-mesospins freeze into a pre-defined disordered sub-lattice, the Shakti ground state manifold is restricted and shaped by the long-mesospin sub-lattice. When coupling four Type-I$_4$ vertices via a long-mesospin, 16 different Type-I$_4$-tilings are possible. The Shakti ground state, as defined by Chern {\it et al.}\cite{Chern2013PRL}, can be obtained with only 12 out of the total 16 tilings. The mSH lattice, with its mono-sized short-mesospins, can access the entire ensemble of tilings. However, the pre-defined disordered long-mesospin arrangement effectively reduces the possible states in the Shakti ground state manifold (see Supplementary Materials Fig.~4). The frozen long-mesospin restricts the Type-I$_4$-tiling amongst four four-fold coordinated vertices to 7 out of 16 possible arrangements. As the system grows in size, the Shakti ground state manifold is further reduced by each added frozen long-mesospin (for further information see Supplementary Materials). This is a plausible cause for the higher order found in the mSH lattice for the strong-coupling regime reported by Gilbert {\it et al.}\cite{Gilbert2014NatPhys} compared to the SH lattice. However, as the inter-island coupling strength is reduced, the long-mesospins, although a source of randomness, provide a better locally-correlated state than the two-fold coordinated vertices of the mSH lattice.

\section{Conclusions}

Using thermally-active nano-patterned arrays of weakly-coupled magnetic mesospins, we have highlighted the impact of distinct length- and energy-scales on the development of magnetic order in artificial spin ice. While our study is restricted to the Shakti geometry and the weak-coupling regime, it captures the importance of the energy hierarchy in mesoscopic architectures and its role for the obtained magnetic order. Given the vast opportunities in geometrical design and the available choices of magnetic materials, the energy hierarchy can be engineered, and, in combination with suitable thermal or even demagnetization protocols, used as a tool for tailoring the magnetic order in mesoscopic magnetic structures. The hierarchy of the energy-scales, as well as the interplay between them, lie behind the complex emergent behavior and properties seen frequently across a range of natural systems\cite{Anderson_Science_1972}. Artificial spin systems provide, therefore, a unique opportunity to study emergence, by offering a `designer' fabrication route and parametrization of the relevant variables. This opportunity potentially allows for a better understanding of problems in physics requiring multiple scale-analysis\cite{Wilson_SciAm_1979}.

\section{Acknowledgments}
The authors would like to thank C.~Nisoli and G.W.~Chern for valuable discussions. The authors acknowledge support from the Knut and Alice Wallenberg Foundation, the Swedish Research Council and the Swedish Foundation for International Cooperation in Research and Higher Education. The patterning was performed at the Center for Functional Nanomaterials, Brookhaven National Laboratory, supported by the U.S. Department of Energy, Office of Basic Energy Sciences, under Contract No. DE-SC0012704. This research used resources of the Advanced Light Source, which is a DOE Office of Science User Facility under contract No. DE-AC02-05CH11231. 
U.B.A. acknowledges funding from the Icelandic Research Fund grants Nr. 141518 and 152483.

\bibliographystyle{natphys}
\bibliography{AllPapers}

\begin{thebibliography}{10}
\bibitem{Incoherent}
 With the term incoherent, we mean that we do not account for the phase
  relationship between the averaged sampling windows. The correlations
  represented in the calculated SSF maps, are only up to lateral extensions of
  the respective sampling window size and are insensitive to correlations
  between or across multiple sampling windows.
\end{thebibliography}

\section{Author contributions}
H.S. D.G. and A.S. fabricated the samples. H.S., E.\"{O}., U.B.A., and V.K. performed the PEEM-XMCD experiments. H.S., E.\"{O}., I.A.C., D.G, T.P.A.H., B.H. and V.K. analyzed the data and contributed to theory development. H.S., I.A.C., B.H. and V.K. wrote the manuscript. All authors discussed the results and commented on the manuscript.

\end{document}


\title{Magnetic order and energy-scale hierarchy in artificial spin ice}

\author{Henry Stopfel}%
\email{henry.stopfel@gmail.com}
\affiliation{%
 Department of Physics and Astronomy, Uppsala University, Box 516, SE-75120, Uppsala, Sweden
}%
\author{Erik \"{O}stman}
\affiliation{%
 Department of Physics and Astronomy, Uppsala University, Box 516, SE-75120, Uppsala, Sweden
}%
\author{Ioan-Augustin Chioar}%
\affiliation{%
 Department of Physics and Astronomy, Uppsala University, Box 516, SE-75120, Uppsala, Sweden
}%
\author{David Greving}
\affiliation{
 Department of Physics, University of Warwick, Coventry, United Kingdom
}%
\author{Unnar~B.~Arnalds}%
\affiliation{%
 Science Institute, University of Iceland, Reykjavik, Iceland
}%
\author{Thomas P. A. Hase}
\affiliation{
 Department of Physics, University of Warwick, Coventry, United Kingdom
}%
\author{Aaron Stein}%
\affiliation{%
 Center for Functional Nanomaterials, Brookhaven National Laboratory, Upton, New York 11973, USA
}%
\author{Bj\"{o}rgvin Hj\"{o}rvarsson}%
\affiliation{%
 Department of Physics and Astronomy, Uppsala University, Box 516, SE-75120, Uppsala, Sweden
}%
\author{Vassilios Kapaklis}%
\affiliation{%
 Department of Physics and Astronomy, Uppsala University, Box 516, SE-75120, Uppsala, Sweden
}%

\date{\today}
\maketitle

\section{Activation energies}

\begin{figure}[b]
\includegraphics[width=\columnwidth]{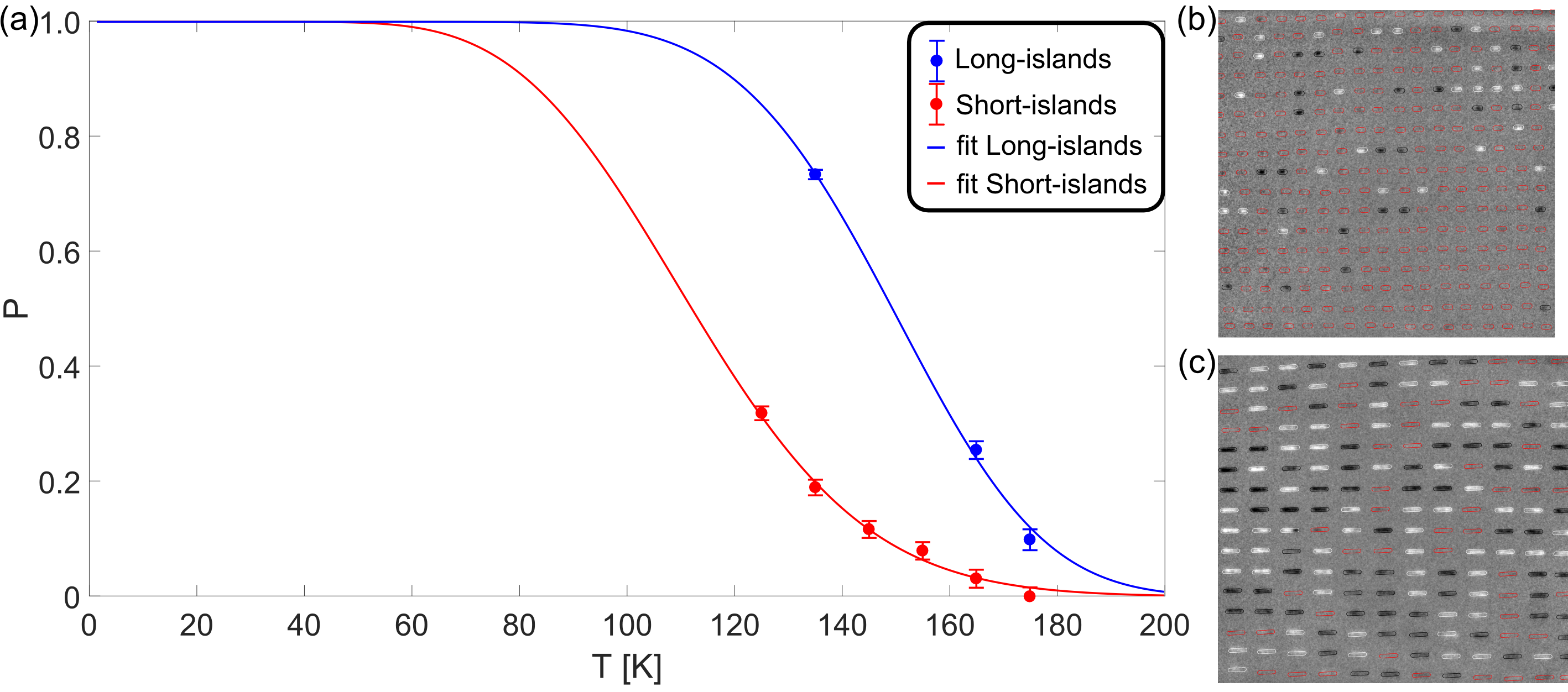}
\caption{Temperature dependent measurements of the thermal dynamics of uncoupled mesospins. ({\bf a}), Fractions of identifiable magnetic islands during a PEEM-XMCD measurement of 300~s and their corresponding fits. Both island types have a width of 150~nm, while the short islands have a length of 450~nm and the long of 1110~nm. Examples of the PEEM-XMCD images for the short ({\bf b}) and long islands ({\bf c}) are presented for the measurement at 135~K.}
\label{activ}
\end{figure}

\noindent To determine the activation energies or the ratio between the activation energies for short- and long-mesospins, we fabricated arrays of short- and long-mesospins with an edge to edge distance of 1000~nm. For these islands this distance is large enough to consider the islands to be non-interacting.
%
%

The fractions of identifiable islands for short (red) and long islands (blue) can be fitted with a double exponential of the following form:
\begin{table*}[h!]
	\centering
		\begin{tabular}{ll}
			\parbox{200px}{\begin{equation}
											P(T) = exp\left(-\dfrac{A}{exp\left(\dfrac{B}{T}\right)}\right)\hspace{20px}\label{eq:fit}
											\end{equation}}& \vspace{-80px}\\
			 & \parbox{200px}{\begin{equation}
													A = \dfrac{t_s}{\tau_0} = \dfrac{300~\rm{s}}{\tau_0}\hspace{20px}\label{eq:A}
												\end{equation}}\vspace{-10px}\\
			 & \parbox{200px}{\begin{equation}
													B = \dfrac{E}{k_B}\hspace{58px}\label{eq:B}
												\end{equation}}
		\end{tabular}
\end{table*}

\vspace{-20px}\noindent Equation \eqref{eq:fit} represents the probability of a magnetic element not having altered its magnetization direction after a time period $t_s$\cite{Wernsdorfer1997PRL}, $T$ is the temperature and $A$, $B$ are fitting parameters. 
The measurement time for the PEEM-XMCD imaging was $t_s = 300$~s. In equation \eqref{eq:A}, $\tau_0$ is the inverse attempt frequency\cite{Kapaklis2014NatNano,Bedanta2009}, while in equation \eqref{eq:B}, $E$ is the activation energy and $k_B$ is Boltzmann's constant. The ratio of the activation energies for short ($E_{short}/k_B = 560(40)$~K) and long islands ($E_{long}/k_B = 1140(140)$~K) can be determined from the fits shown in Suppl. Fig. \ref{activ}({\bf a}), and is 0.49(4).
\section{Real Space Images}
\begin{figure}[h!]
\includegraphics[width=\columnwidth]{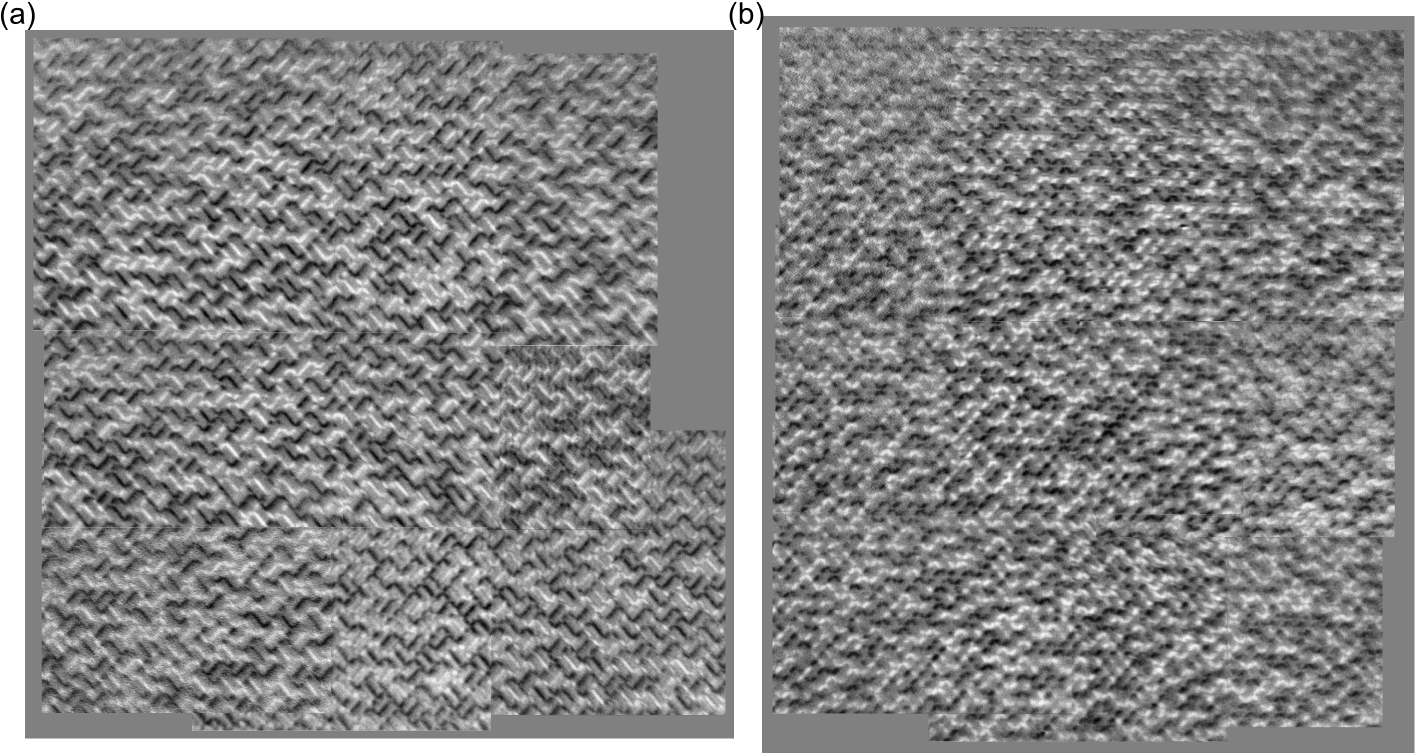}
\caption{PEEM-XMCD measurements of the Shakti lattices. Composite PEEM-XMCD images of the SH ({\bf a}) and mSH ({\bf b}) lattices.}
\label{PEEM}
\end{figure}

Real space images of the SH and mSH lattices that are created by stitching together individual images (each recorded for 360~s) recorded in the limited field of view of the microscope. The composite images shown in Fig.~\ref{PEEM}({\bf a}) for the SH and Fig.~\ref{PEEM}({\bf b}) for the mSH lattices represent the raw data from which the vertex statistics and spin structure factors were calculated. The individual islands can be seen as either black/white/gray depending on whether the islands are aligned parallel/anti-parallel/reversing with respect to the x-ray propagation direction. The particular composite images shown in Fig.~\ref{PEEM} were recorded at 65~K.

\newpage
\section{Spin arrangements}

\begin{figure}[b]
\includegraphics[width=0.97\columnwidth]{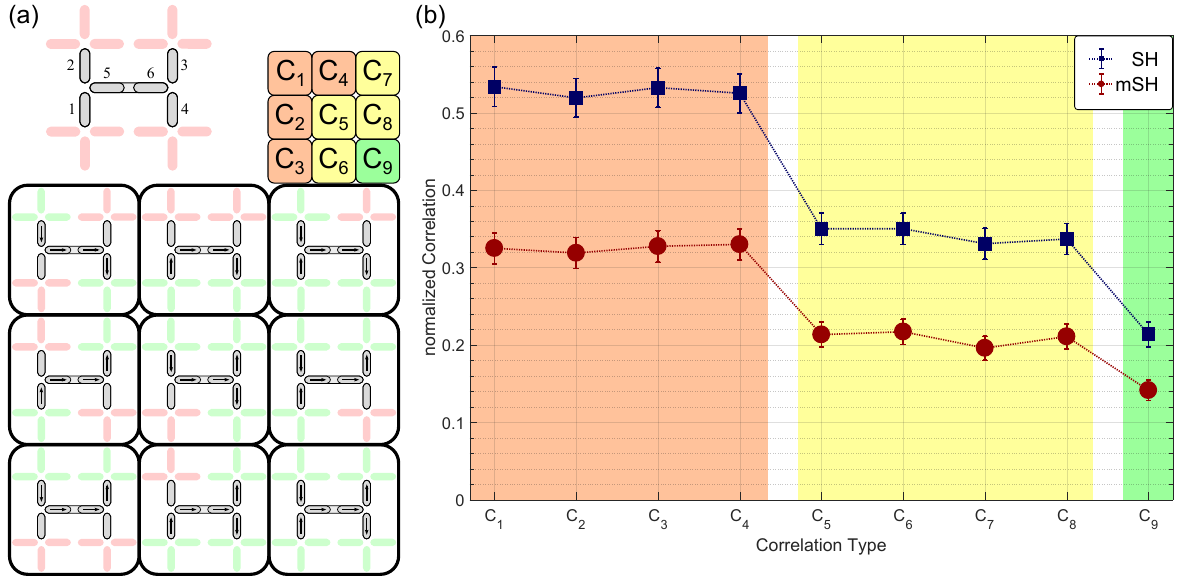}
\caption{Spin arrangement arround the long-mesospin (Type-I$_2$). ({\bf a}), Illustration of the five (six in mSH) islands which are considered in the spin arrangements. Adjacent islands are indicated by light red and in the different representation of the spin arragements C$_1$-C$_9$ with light green if correlated over the long-mesospin (Type-I$_2$). ({\bf b}), Abundance of the different spin arrangements, for the SH (dark-blue -- squares) and mSH (dark-red -- circles) lattices.}
\label{Shakticorrelation}
\end{figure}

All nine possible spin arrangements $\rm{C}_1-\rm{C}_9$, for the short- and long-mesospins (depending on the lattice) of two neighboring three-fold coordinated vertices are depicted in Fig. \ref{Shakticorrelation}({\bf a}). The abundance of the spin arrangements is presented in Fig. \ref{Shakticorrelation}({\bf b}). These spin arrangements can be categorized into three groups marked with orange, yellow and green. Group one (orange~--~A$_1$), are arrangements $\rm{C}_1-\rm{C}_4$ connecting one mesospin, and the respective vertex, with another mesospin on the other side of the long-mesospin (two short-mesospins). The second group (yellow~--~A$_2$), are arrangements $\rm{C}_5-\rm{C}_8$ in which one mesospin on one side is in ferromagnetic alignment with the long-mesospin (two short-mesospins) and two mesospins on the other side. Finally, the last arrangement (green), $\rm{C}_9$, is the lowest energy state for this sub-set of mesospins.

Roughly 53\% of all investigated long-mesospin connections of the SH have at least two vertices being correlated to each other ($\rm{C}_1-\rm{C}_4$), but only 32\% of them are correlated in the mSH via the two short-islands. Similar differences in the correlations can also be found for the other spin arrangements ($\rm{C}_5-\rm{C}_9$). This can be attributed to the high abundance (34\%) of Type-II$_2$ excitations in vertices with coordination number two, which are therefore unable to mimic the SH order.

\section{Type-I$_4$-tiling}
\begin{figure}[h]\vspace{-10px}
\includegraphics[width=\columnwidth]{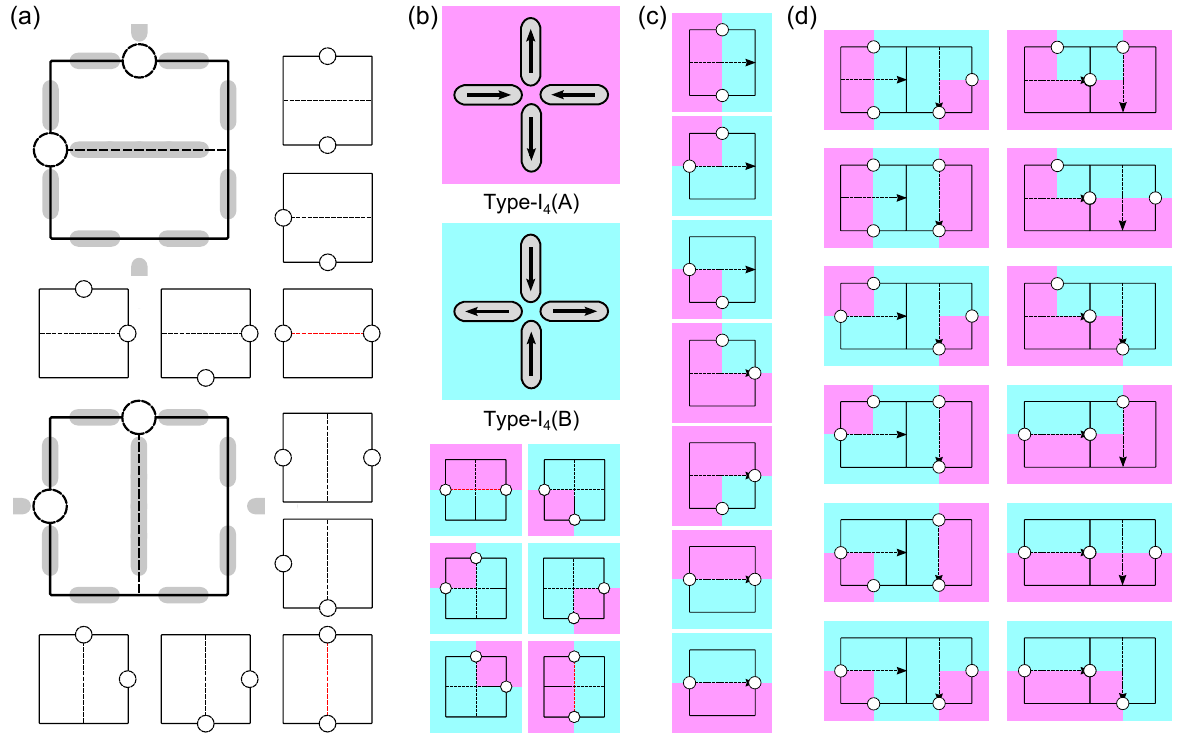}
\caption{Shakti ground state manifold in plaquettes. ({\bf a}), Illustration of the Shakti ground state manifold in a plaquette as introduced by Chern {\it et al.}\cite{Chern2013PRL}. In the Shakti ground state all four-fold coordinated vertices are in their lowest energy state, Type-I$_4$, and 50\% of the three-fold coordinated vertices are in their lowest energy state, Type-I$_3$, and 50\% are in the excited state, Type-II$_3$. These excitations are represented with a circle in the plaquette illustration by Chern {\it et al.}. ({\bf b}), Equivalent representation of the Shakti ground state manifold for a plaquette, focusing on the Type-I$_4$ tiling at the four-fold coordinated vertices. All Type-I$_4$ tilings for a pre-defined long-mesospin direction are illustrated for one ({\bf c}) and two  plaquettes ({\bf d}).}
\label{sup3}
\end{figure}
The ground state manifold of the Shakti lattice was described by Chern {\it et al.}\cite{Chern2013PRL} and is best explained on the size of a plaquette. In Fig.~\ref{sup3}({\bf a}) the extent of a plaquette is illustrated, consisting of 9 islands in the SH and 10 islands in the mSH case. 
A Shakti plaquette has one long island (or a two-fold coordinated vertex in the mSH) in its center and 8 short islands at its borders, sharing 4 three-fold coordinated vertices and 4 four-fold coordinated vertices with the neighbouring plaquettes. 
Fig.~\ref{sup3}({\bf a}) also shows the representation of the Shakti ground state manifold of such a plaquette as introduced by Chern {\it et al.}\cite{Chern2013PRL}. Solid lines illustrate the lowest energy states, but the circles at the three-fold coordinated vertices mark the position of Type-II$_3$ excitations. In the Shakti ground state a plaquette inhabits two of such Type-II$_3$ excitations, at any of their 4 three-fold coordinated vertices. At the same time all the four-fold coordinated vertices are in their lowest energy state, Type-I$_4$. Therefore we represent the tiling of the two degenerated Type-I$_4$(A) and Type-I$_4$(B) vertices on the bases of a plaquette in Fig~\ref{sup3}({\bf b}).

The excitations in the Shakti lattice are located at the boundaries between Type-I$_4$(A) and Type-I$_4$(B) vertices. We discussed in the main text, that the pre-defined long-mesospins in the SH lattice will restrict the possible Type-I$_4$-tilings. This effect is illustrated in Fig.~\ref{sup3}({\bf c})-({\bf d}) for one and two plaquettes. All 7 (out of 16) Type-I$_4$-tilings accessible, by simultaneously remaining in the Shakti ground state manifold, regarding the given long-mesospin direction are illustrated in Fig.~\ref{sup3}({\bf c}). For two pre-defined long-mesospins (two neighbouring plaquettes -- Fig.~\ref{sup3}({\bf d}) ) just 12 (out of 64) Type-I$_4$-tilings remain in the Shakti ground state manifold.

\bibliographystyle{natphys}
\bibliography{AllPapers_2}